# On-fiber high-resolution photonic nanojets via high refractive index dielectrics

**WASEM ALJUAID,**[1] **JOSEPH ARNOLD RILEY,**[1,2] **NOEL HEALY,**[1] **AND VICTOR PACHECO-PEÑA,**[1,*]

[1] *School of Mathematics, Statistics and Physics, Newcastle University, Newcastle Upon Tyne NE1 7RU, United Kingdom*
[2] *School of Engineering, Newcastle University, Newcastle Upon Tyne NE1 7RU, United Kingdom*
*victor.pacheco-pena@newcastle.ac.uk*

**Abstract:** In this manuscript, we present high spatial resolution focusing of electromagnetic waves at telecommunication wavelengths ($\lambda_0$ = 1.55 μm) by using high-refractive index mesoscale dielectrics placed at the end of an optical fiber. Our approach exploits photonic nanojets (PNJs) to achieve high-intensity, spatially narrow focal spots. The response of the device is evaluated in detail considering 2-dimensional (2D) and 3-dimensional (3D) configurations using high-index mesoscale cylindrical and spherical dielectrics, respectively, placed on top of an optical fiber. It is shown how the PNJs can be shifted towards the output surface of the mesoscale high-index dielectric by simply truncating its 2D/3D cylindrical/spherical output profile. With this setup, a PNJ with a high transversal resolution is obtained using the 2D/3D engineered mesoscale dielectric particles achieving a Full-Width at Half-Maximum of *FWHM* = 0.28$\lambda_0$ (2D truncated dielectric), and $FWHM_y$ = 0.17$\lambda_0$ and $FWHM_x$ = 0.21$\lambda_0$ (3D truncated dielectric). The proposed structure may have potential in applications where near-field high spatial resolution is required, such as in sensing and imaging systems.

## 1. Introduction

Detection of subwavelength particles has long been a challenging proposition due to the intrinsic limitations imposed by the natural diffraction of electromagnetic waves, such as the well-known Abbe barrier [1–3]. To overcome such a diffraction limit, the scientific community has proposed and demonstrated different ways to enhance the spatial resolution of focal spots produced by lenses for imaging and sensing applications. Proposed techniques include superoscillatory lenses [4–6], superlenses [7–11], and micro/nanometer-scale dielectrics [3,12–18], to name a few. Among these methods, photonic nanojets (PNJs) have been demonstrated to improve the spatial resolution of focusing structures enabling the detection of subwavelength objects [15,17,19,20].

PNJs are high-intensity spatial spots produced at the shadow (output) side of a dielectric particle (with varying shapes such as spheres, cylinders, cuboids, and non-symmetrical dielectrics [21–25]) when illuminated with a planewave. In this context, a pioneering work was proposed several years ago when cylindrical or spherical dielectrics were illuminated by a planewave to produce a focal spot with a spatial resolution below the diffraction limit [19,26]. From this, a great deal of research has been done to find different methods to generate and improve the performance of PNJs using multiple configurations such as PNJs generated using waveguides [27–29] and also exploiting graded-index and core-shell techniques [30,31]. They have been implemented in a vast frequency range, from acoustics and microwaves up to the optical regime [32,33], showing how they can also be exploited in compact plasmonic systems using surface plasmon polaritons [34–37]. Several applications have benefited from the introduction of PNJs such as microscopy and imaging of subwavelength particles [13,38,39], Raman spectroscopy [40], sensing of a single molecule [41], and surgery in medicine [42].

As it is known in the literature, to produce PNJs at the output surface of mesoscale dielectric particles illuminated with a planewave, the ratio between the refractive index of the dielectric, $n_d$, and the surrounding background medium, $n_b$, should be less than two (i.e., $n_d/n_b < 2$) [34,35,43–48]. Recent efforts to overcome this include the use of Gaussian beams to illuminate high-index spheres [49]. Recently, we have reported how such conditions can be relaxed when exploiting hemispherical mesoscale high-index dielectric particles [21]. We have shown how techniques borrowed from solid immersion lenses (such as the Weierstrass formulation [50]) can be exploited to truncate the profile of mesoscale high-index cylindrical (2D) or spherical (3D) dielectrics. In this realm, when illuminated by a planewave, such structures can indeed produce high intensity PNJs near their output surface, opening further opportunities for PNJs in applications that require a high spatial resolution. Moreover, as high-index 2D/3D truncated cylinders/spheres are implemented ($n_d/n_b > 2$), their size is reduced compared to PNJs generated by cylinders/spheres that hold the ratio $n_d/n_b < 2$ [51,52], allowing the design of potentially more compact devices, as expected. However, despite the great opportunities offered by PNJs using dielectric particles, one of the challenges is the fact that they are designed to be free-standing with air as the background medium, making their application in real scenarios a challenging task.

Inspired by the interesting properties of PNJs, their ability to be excited using compact high-index mesoscale dielectrics and the need to overcome free-standing designs, in this work we build upon our recent findings [21,53] to design and study a structure with the ability to generate PNJs at telecommunication wavelengths using truncated high refractive index mesoscale dielectrics cylinders/spheres positioned on top of an optical fiber. The design process of the structure is presented, and their response is evaluated considering idealized 2D cylinders and realistic 3D spheres. It is discussed how, similar to our previous findings [21] the location of the PNJ can be shifted near the output surface of the 2D/3D structures by truncating the cylindrical/spherical profiles. The influence of dielectric losses is presented in detail along with designs considering 3D dielectric structures being not fully spherical but flat on their input surface to account for potential fabrication imperfections and to relax the need of full hemispherical profiles. As the dielectric structure is placed on top of the optical fiber, the proposed device (optical fiber + dielectric particle) may enable free control of the location of the PNJ via mechanical or manual positioning of the whole structure, relaxing the need for free-standing designs [20,27–29,54]. The proposed device may find applications in scenarios where high spatial resolution is needed such as in imaging and sensing systems at telecommunication wavelengths.

## 2. Methods and materials

To begin with, we make use of a single-mode optical fiber consisting of a cylindrical core and cladding working at telecom wavelengths ($\lambda_0$ = 1.55 μm). The importance of using a single-mode optical fiber is to remove all of the higher modes inside the fiber [55] which could deteriorate the overall response of the generated PNJ. For instance, for PNJs generated via tip-ended optical fibers, it has been shown that higher modes can lead to the maximum intensity outside of the fiber being shifted off-axis [28]. In this context, for our single mode optical fiber, silicon dioxide slightly doped with germanium ($SiO_2$-Ge) is used as the material for the core ($n_{co}$ = 1.445) while silicon dioxide ($SiO_2$) is then used for the cladding ($n_{clad}$ = 1.44). The core and cladding have a diameter of 8 μm and 125 μm, respectively. Since the aim of the proposed structure is to obtain a PNJ with a high spatial resolution, a truncated mesoscale high refractive index cylinder/sphere is implemented on top of the $SiO_2$-Ge optical fiber. For this mesoscale dielectric, and without loss of generality, we chose a refractive index of $n_d$ = 3.3 which is near to the value of commonly used materials at telecom wavelengths such as silicon [1,56]. Finally, air is used as the surrounding medium (background $n_b = 1$). In this context, in line with [21] the ratio between the dielectric and the background media will be $n_d/n_b > 2$.

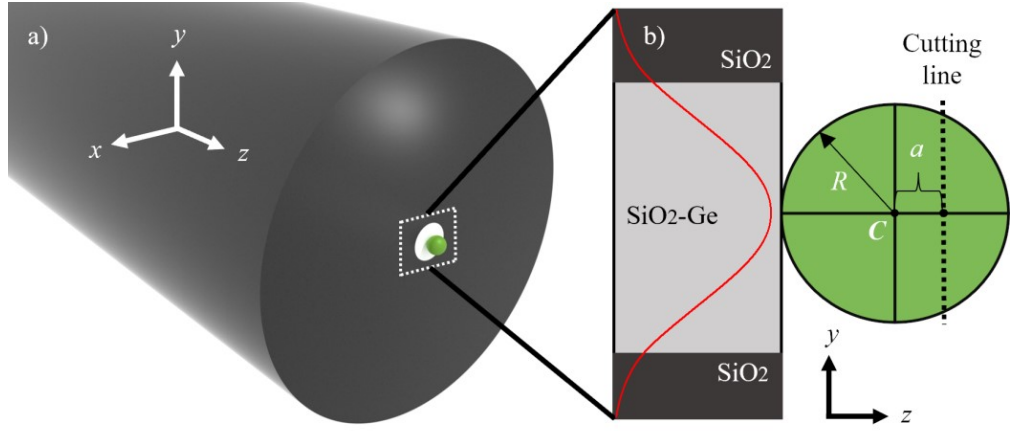

Fig. 1. (a) Schematic representation of the optical fiber with the mesoscale dielectric particle placed at its end and immersed in air. (b) Side view of (a). The red line represents the power distribution of the fundamental mode inside the fiber. The design parameters for the truncated mesoscale dielectric structure are also shown in (b).

A schematic representation of the proposed structure is presented in Fig. 1a where the single-mode optical fiber with a mesoscale spherical dielectric placed at the end of it is shown. For completeness, a side view is also shown in Fig. 1b along with the profile of the fundamental mode of the fiber as a red curve. From now on, we will consider that the optical fiber is illuminated with the electric field polarized along the $y$-axis ($E_y$ polarization) with propagation along the $z$-axis. Due to the high refractive index of the mesoscale dielectric sphere with respect to air ($n_d/n_b > 2$), one would expect the PNJ to be generated inside of the high index dielectric [21]. To overcome this, we can exploit the process outlined in our recent work [21] and truncate the high index mesoscale dielectric particle using the Weierstrass formulation for immersion lenses shown in Eq. 1 [50].

$$a = \left(1 + \frac{1}{n_d}\right)R - R \qquad (1)$$

where $a$ is the truncated distance from the center ($C$) of the dielectric and $R$ is the radius of the mesoscale dielectric particle (see Fig. 1b). With this configuration, the mesoscale dielectric is cut along the dotted line represented in Fig. 1b. It is important to note that Eq. 1 is valid for immersion lenses surrounded by air when illuminated with a planewave. In our work, however, we make use of the fundamental mode of a single mode optical fiber (as in Fig. 1). In this realm, while we make use of Eq. 1 as a starting point, the final truncation distance $a$ will need to be optimized, as it will be shown in sections 3 and 4 below.

## 3. High-index mesoscale cylinder dielectric (2D)

To evaluate the performance of the proposed structure, a numerical study is first carried out (using the commercial software COMSOL Multiphysics®) considering an idealized 2-dimensional (2D) setup with the dielectric positioned at the output face (right) of the fiber. The structure (2D fiber and dielectric) is immersed in air with top, bottom, and right scattering boundary conditions to avoid undesirable reflections. The 2D fiber was illuminated from the left boundary using a port covering the whole fiber (core and cladding) with the electric field polarized along the vertical $y$-axis propagating along the positive $z$-axis. Finally, an extremely fine mesh was used with a maximum and minimum mesh size of $0.1\lambda_0$ and $6\times10^{-4}\lambda_0$, respectively. An additional automatic refinement was applied to further improve the mesh. With this setup, a full 2D dielectric cylinder is first considered to evaluate its performance prior to applying the truncation method (described in the last section). To do this, different sizes of the 2D lens are analyzed ranging from $R = 0.55\lambda_0$ up to $R = 1.55\lambda_0$ (to be in line with our previous findings using cylinders and spheres in free-space [21]) with steps of $0.1\lambda_0$. We focus our study on the spatial resolution, position of the focus (focal length, *FL*), and power enhancement (the ratio between the power distribution at the focal plane with and without using the high-index mesoscale dielectric). To evaluate the spatial resolution of the generated focus, the Full-Width at Half-Maximum (*FWHM*, distance along the transversal axis at which the power distribution has decayed half of its maximum) is calculated [31,57].

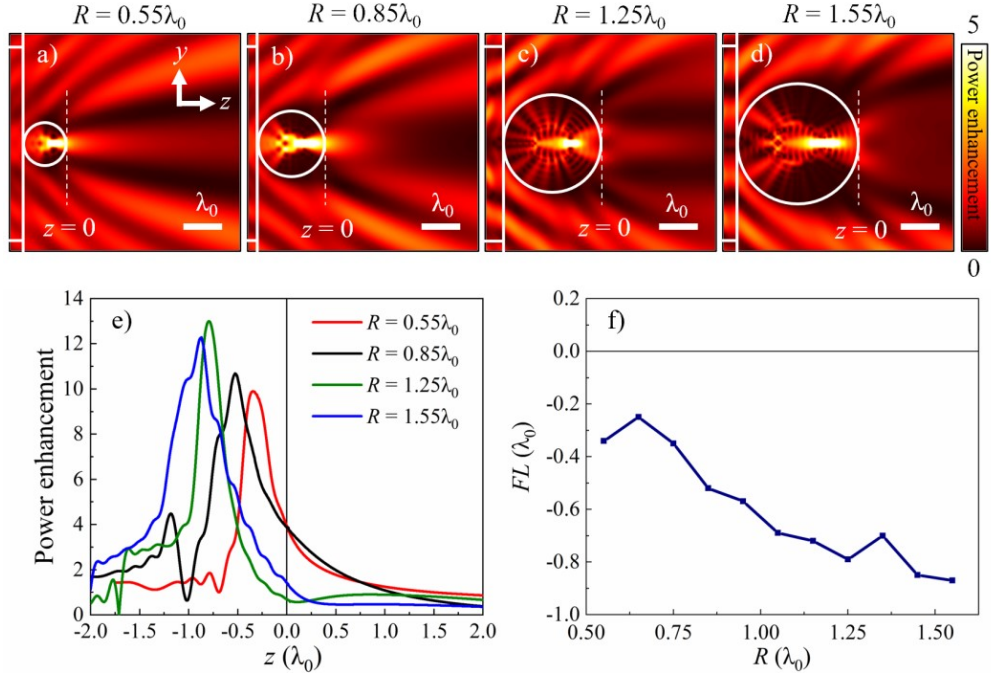

Fig. 2. Power enhancement on the $yz$-plane considering a radius $R$ of the full 2D dielectric cylinder of (a) $R = 0.55\lambda_0$, (b) $R = 0.85\lambda_0$, (c) $R = 1.25\lambda_0$, and (d) $R = 1.55\lambda_0$. Note that the contour scales are saturated from 0 to 5 to better compare the results. (e) Power enhancement along the $z$-axis at $y = 0$ (extracted from panels a-d) for the same values of $R$ as in a-d. (f) Focal length (*FL*) for all values of $R$ from $R = 0.55\lambda_0$ to $R = 1.55\lambda_0$ with steps of $0.1\lambda_0$. The vertical and horizontal black lines in (e) and (f) represent the shadow face of the full 2D dielectric cylinders.

With this setup, the numerical results of the power enhancement on the $yz$-plane for the different values of $R$ of the full 2D dielectric cylinders are shown in Fig. 2a-d, respectively. As observed, the generated focus is located inside of the cylinders. This is an expected result due to the relatively high value of $n_d$ with respect to $n_b$ (i.e., $n_d/n_b = 3.3$, larger than 2) [21,47]. To better compare the results, the power enhancement along the $z$-axis at $y = 0$ was extracted from Fig. 2a-d and it is plotted in Fig. 2e. From here, it is evident how the focus is produced inside the cylinders (the vertical black line in this figure represents the shadow surface of the cylinders, located at $z = 0$, and positive $z$ is free-space). The values of the *FL* of the generated focal spots are shown in Fig. 2f as a function of $R$ of the cylinders. To guide the eye, the shadow surface of the mesoscale dielectric cylinders is also represented as the horizontal black line at zero in this panel. As it is shown, the *FL* is moved far from the output surface of the dielectrics (left direction in Fig. 2e) as $R$ increases. Based on this, it is clear how increasing $R$ shifts the *FL* further inside of the cylinders given that $n_d/n_b > 2$ in agreement with our recent finding [21], which is the opposite of what occurs when exciting PNJs with $n_d/n_b < 2$ [19,26] (where increasing $R$ moves the PNJ from inside the dielectric towards free-space). In terms of the spatial resolution, the *FWHM* along the transversal *y*-axis was calculated at the corresponding *FL* for each value of $R$ with values of $0.46\lambda_d$, $0.56\lambda_d$, $0.42\lambda_d$, and $0.46\lambda_d$ for $R = 0.55\lambda_0$, $0.85\lambda_0$, $1.25\lambda_0$, and $1.55\lambda_0$, respectively (with $\lambda_d \approx 470$ nm as the wavelength inside the dielectric cylinders).

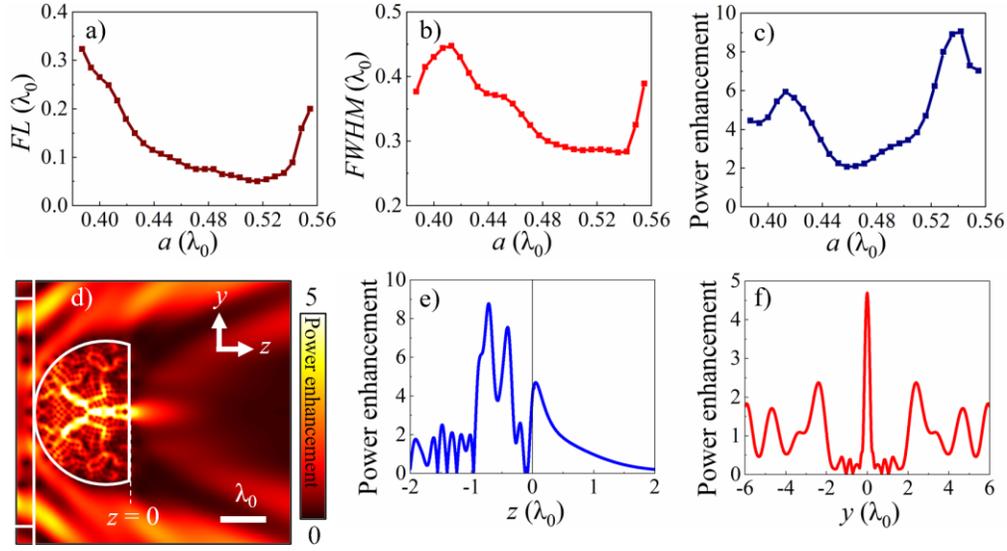

Fig. 3. Numerical results of the (a) *FL*, (b) *FWHM* and (c) power enhancement as a function of $a$ for a truncated 2D dielectric cylinder with $R = 1.55\lambda_0$. (d) Power enhancement on the *yz*-plane for $R = 1.55\lambda_0$ and $a \approx 0.52\lambda_0$. Power enhancement along (e) the *z*-axis at $y = 0$ and (f) *y*-axis at the *FL*, extracted from panel (d).

To move the generated PNJ near the shadow surface of the mesoscale dielectric, we can truncate the cylinder as described in section 2. To do this, we choose the radius $R$ of the cylinder as $R = 1.55\lambda_0$ (to be of a similar size, in terms of the wavelength inside the dielectric cylinder, to our previous works under planewave illumination [21]). We truncated the cylinder from Fig. 2d (with $R = 1.55\lambda_0$) in order to move the PNJ near the shadow surface of the dielectric using Eq. 1. In this case, the resulting value is $a \approx 0.47\lambda_0$. Importantly, as we consider that the dielectric is on top of an optical fiber, the illumination is different than a planewave (as considered in [21]), therefore, it is expected that $a$ will need to be slightly tuned to account for such change of illumination. To do this, a parametric study of the performance of the structure is carried out considering values of $a$ (for the fixed $R = 1.55\lambda_0$) ranging from $\sim 0.39\lambda_0$ to $\sim 0.55\lambda_0$ with steps of $\sim 0.006\lambda_0$ (600 nm to 860 nm with a step of 10 nm).

With this configuration, the numerical results of the *FL, FWHM*, and power enhancement as a function of the truncated distance, $a$, are shown in Fig. 3a-c, respectively. From Fig. 3a, it can be seen how the foci appear beyond the shadow surface of the truncated 2D dielectric (with a value of zero being equals to the position of the shadow flat surface of the dielectric). Moreover, it can be observed that the foci are moved toward the flat surface of the mesoscale dielectric as the value of $a$ increases, as expected (see discussion above regarding the ratio $n_d/n_b > 2$). The closest position of the focus to the output flat surface is achieved for $a \approx 0.52\lambda_0$ where the focus is $0.05\lambda_0$ away from the output surface (outside the dielectric). By looking at the spatial resolution of the generated PNJs (Fig. 3b) all of the *FWHM* values are smaller than $\lambda_0/2$. Moreover, note how the resolution is improved for those PNJs near the shadow surface, in agreement with [26], with the smallest *FWHM* value ~$0.28\lambda_0$ when *a* is between $0.50\lambda_0$ to $0.54\lambda_0$. Finally, from the power enhancement shown in Fig. 3c, it is observed how the largest values are obtained when *a* is further increased up to *a* ~$0.55\lambda_0$. However, the spatial resolution is slightly deteriorated given that the location of the generated PNJs is shifted away from the output surface of the 2D truncated dielectric, as expected [21,22,30].

Based on the results from Fig. 3a-c, the best design can be selected considering the value of *a* that allows for the excitation of a PNJ with the smallest *FWHM* and the closest *FL* to the output surface. As described above, in this case with $R = 1.55\lambda_0$, the PNJs with the smallest resolution are those produced by the truncated 2D dielectric cylinders with $a \approx 0.50\lambda_0$ to $0.54\lambda_0$ while the closest *FL* to the output surface is the design with $a \approx 0.52\lambda_0$. By considering these conditions, for $R = 1.55\lambda_0$, a value of $a \approx 0.52\lambda_0$ can be considered as the best design. This value of *a* is near but not exactly the same as the one calculated with Eq. 1, $a \approx 0.47\lambda_0$, as expected, given that the illumination is no longer an ideal planewave. Note that different values of *a* will be needed if a different value of *R* is chosen, as expected by Eq. 1. For completeness, the results of the power enhancement on the *yz*-plane for the truncated mesoscale dielectric with $a \approx 0.52\lambda_0$ is shown in Fig. 3d-f together with the power enhancement along the propagation *z*-axis (at *y* = 0) and along the *y*-axis at the *FL* ($z = 0.05\lambda_0$), respectively. From these results, a transversal resolution of $FWHM = 0.28\lambda_0$ is achieved with a power at the *FL* ~4 times higher than the case without using the truncated dielectric, demonstrating the ability to produce high spatial resolution PNJs.

## 4. High-index mesoscale dielectric sphere (3D configuration)

Here we investigate the proposed structure in a 3D configuration via the commercial software CST Studio Suite®. For this, the mesoscale dielectric sphere ($R = 1.55\lambda_0$) was truncated using the same values of *a* as in section 3 (see schematic representation in Fig. 4d). The single mode optical fiber was excited using a waveguide port placed at the left side boundary with the electric ($E_y$) and magnetic ($H_x$) fields polarized along the *y*- and *x*-axes, respectively, with propagation along the *z*-axis. The dielectric particle is placed on the right end of the optical fiber and the whole structure (optical fiber and dielectric) was immersed in vacuum. A refined hexahedral mesh was applied with a maximum and minimum mesh size of $0.16\lambda_0$ and $0.016\lambda_0$, respectively. Finally, to reduce the simulation time, electric and magnetic symmetry planes were applied on the *xz*- and *yz*-planes, respectively. With this setup, the numerical results of the generated *FL* as a function of the truncated distance *a* are shown in Fig. 4a. As observed, the PNJs are obtained inside the mesoscale 3D truncated dielectric sphere, unlike the ideal 2D scenario discussed in the previous section. Additionally, by comparing the results from Fig. 4a and Fig. 3a, it can be observed how the PNJs in the 3D configuration are closer (but inside) to the output surface than those obtained with the 2D versions (between ~$|0.02\lambda_0|$ and ~$|0.06\lambda_0|$ from the output surface for the 3D case). This is in line with our previous work under planewave illumination [21] where it was shown how the PNJs were produced inside the dielectrics but were able to leak out of the mesoscale dielectric to free-space.

To evaluate the spatial resolution of the generated PNJs, the *FWHM* along *x*- and *y*-axis have been calculated at the output flat surface, $z = 0$, of the mesoscale 3D truncated dielectric spheres, and the results are shown in Fig. 4b as blue and red lines, respectively. As it is shown, the spatial resolution for all the designs falls below half wavelength in both transversal directions. The highest spatial resolution in *x* and *y* is achieved for $a \approx 0.41\lambda_0$ with $FWHM_x = 0.21\lambda_0$ and $FWHM_y = 0.17\lambda_0$. In addition to the spatial resolution, the power enhancement for all the designs with different values of *a* was measured at the output face of the particle, $z = 0$, and the results are shown in Fig. 4c. As observed, given the 3D nature of the design compared to that of the previous section (2D case), higher power enhancement of the PNJs is obtained with values ranging from ~20 to 70.

Based on the results from Fig. 4b considering the design that produces the smallest *FWHM* along the *x*- and *y*-axes, the numerical results of the power enhancement on the *xz*-plane (*y* = 0) and the *yz*-plane (*x* = 0) for a 3D truncated mesoscale dielectric with $a \approx 0.41\lambda_0$ are shown in Fig. 4e,f, respectively. As it is shown, the PNJ is generated inside the truncated dielectric but leaks out towards free-space. To better appreciate this performance, the power enhancement along the *z*-axis at *x* = *y* = 0 was extracted from Fig. 4e,f and the results are plotted in Fig. 4g. From this figure, the PNJ is positioned at $z = -0.04\lambda_0$ and decays through the output surface of the particle into free-space at *z* = 0 with a power enhancement ~27 at this latter position. For completeness, the spatial resolution of the generated PNJ is shown in Fig. 4h,i where the power enhancement along the transversal *x*- and *y*-axes at *z* = 0 (extracted from Fig. 4e,f) is shown respectively. These results demonstrate the ability of the proposed structure to generate high intensity and high spatial resolution PNJs at the output surface of the truncated 3D high index dielectric sphere placed on top of an optical fiber.

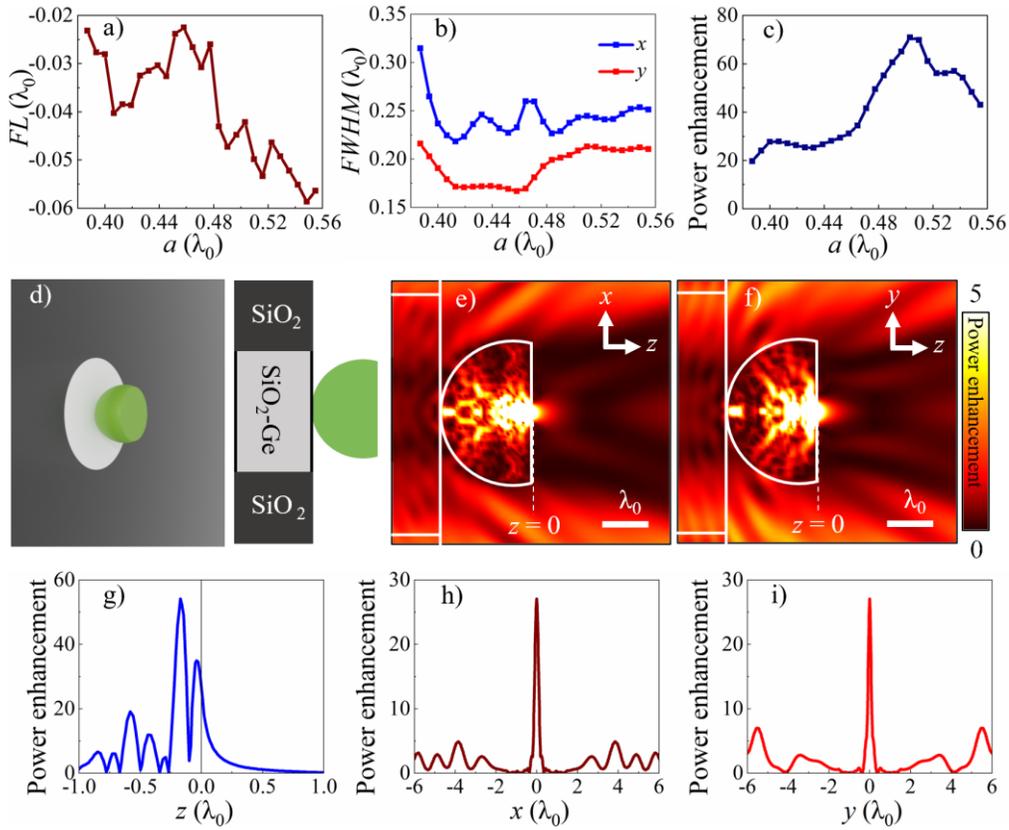

Fig. 4. (a) Position of the *FL* for different values of *a* inside the 3D mesoscale truncated dielectric sphere, where the output surface is set at *z* = 0. (b,c) *FWHM* and power enhancement as a function of *a* calculated at *z* = 0, respectively, for 3D truncated mesoscale dielectric spheres with a radius of $R = 1.55\lambda_0$. (d) Schematic representation of the truncated 3D mesoscale dielectric sphere with $a \approx 0.41\lambda_0$. (e,f) Power enhancement on the *xz*- and *yz*-planes, respectively, for the dielectric with $R = 1.55\lambda_0$ and $a \approx 0.41\lambda_0$. To better display and compare the results, the contour scales have been saturated from 0 to 5. (g) Power enhancement along *z*-axis at *y* = *x* = 0 extracted from the design shown in panels (e,f). The black line at *z* = 0 represents the output surface of the truncated mesoscale dielectric. (h,i) Power enhancement along the *x*- and *y*-axes (calculated at *z* = 0), respectively, extracted from (e) and (f), respectively.

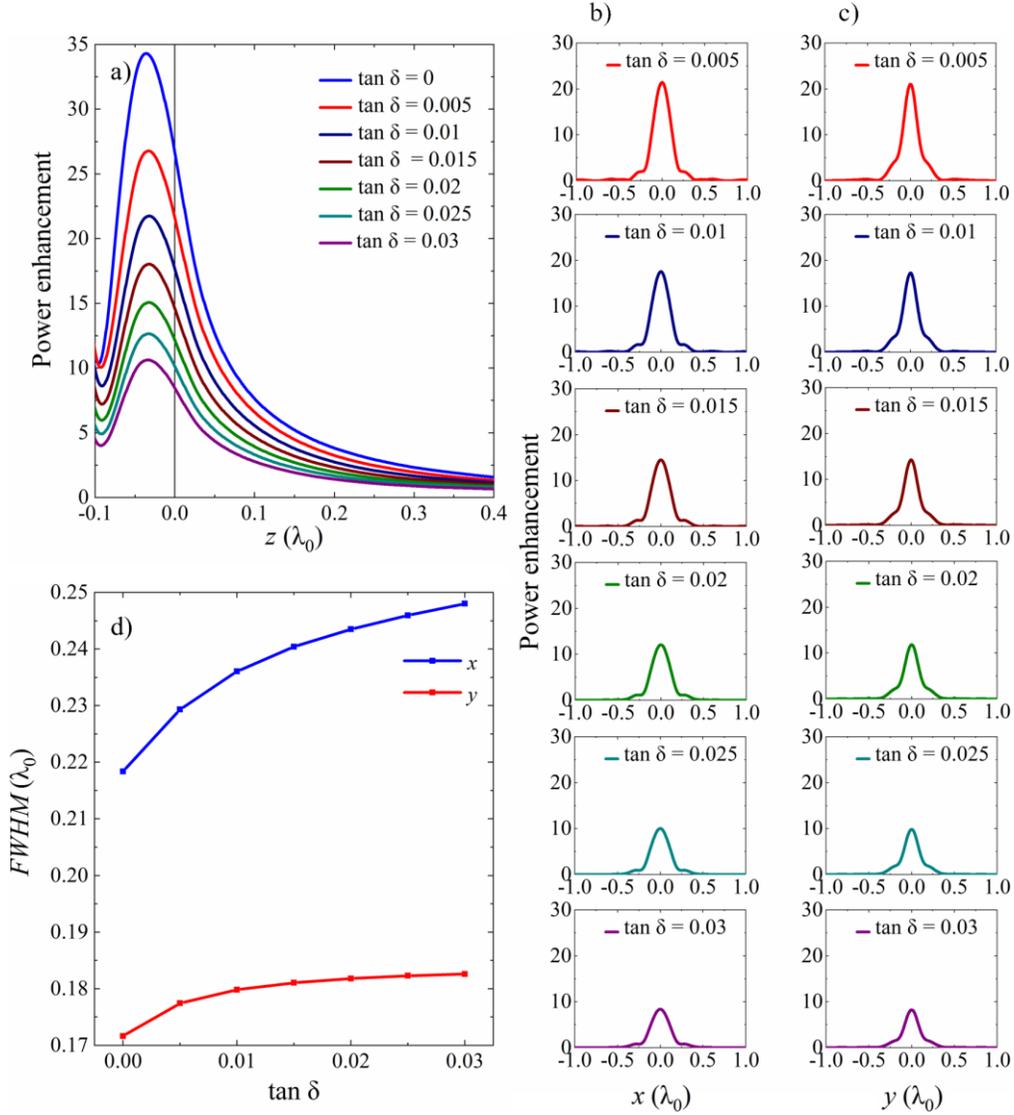

Fig. 5. (a) Power enhancement along $z$-axis at $x = y = 0$ for a truncated 3D dielectric sphere as the one discussed in Fig. 4g when considering dielectric losses with tan δ values ranging from 0 to 0.03 in steps of 0.005. (b,c) Power enhancement along the transverse $x$- and $y$-axes at $z = 0$ respectively for the different values of tan δ. (d) $FWHM_x$ (blue) and $FWHM_y$ (red) as a function of different values of tan δ calculated at $z = 0$ for the truncated mesoscale dielectric.

Now, the results discussed in Fig. 2-4 were calculated considering the truncated mesoscale dielectrics to have negligible losses. For completeness, we discuss here the effect of dielectric losses on the position of the $FL$ of the generated PNJ and its respective spatial resolution as this is an important aspect from an experimental point of view. To do this, we can consider the design discussed in Fig. 4g-i using a 3D truncated dielectric sphere with $a \approx 0.41\lambda_0$ and $R = 1.55\lambda_0$. The numerical results are shown in Fig. 5 when introducing dielectric losses with values of tan δ ranging from 0.005 to 0.03 with steps of 0.005. As shown in the Fig. 5a, as the tan δ is increased, the power enhancement along the propagation $z$-axis of the generated PNJ is reduced, as expected, and its location slightly varies (~0.001$\lambda_0$) within the full range of considered values of tan δ. The effect of introducing losses on the spatial resolution of the generated PNJs can be studied by observing the results from Fig. 5b,c where the power enhancement along the *x*- and *y*-axes (calculated at *z* = 0) are shown, respectively, considering the different values of tan δ under study. From these results it is clear how the power enhancement is reduced, as described before. Moreover, we extracted the values of $FWHM_{x,y}$ from Fig. 5b,c, and the results are presented in Fig. 5d as blue and red lines, respectively. As observed, increasing the dielectric losses reduces the resolution of the generated PNJ to $FWHM_x \approx 0.25\lambda_0$ and to $FWHM_y \approx 0.18\lambda_0$ when considering high losses of tan δ = 0.03. However, even when introducing large dielectric losses, the overall spatial resolution of the generated PNJs remain below ∼$0.25\lambda_0$ along both transversal directions, demonstrating the potential of using the proposed structure to produce PNJs using high index truncated dielectric particles on top of optical fibers.

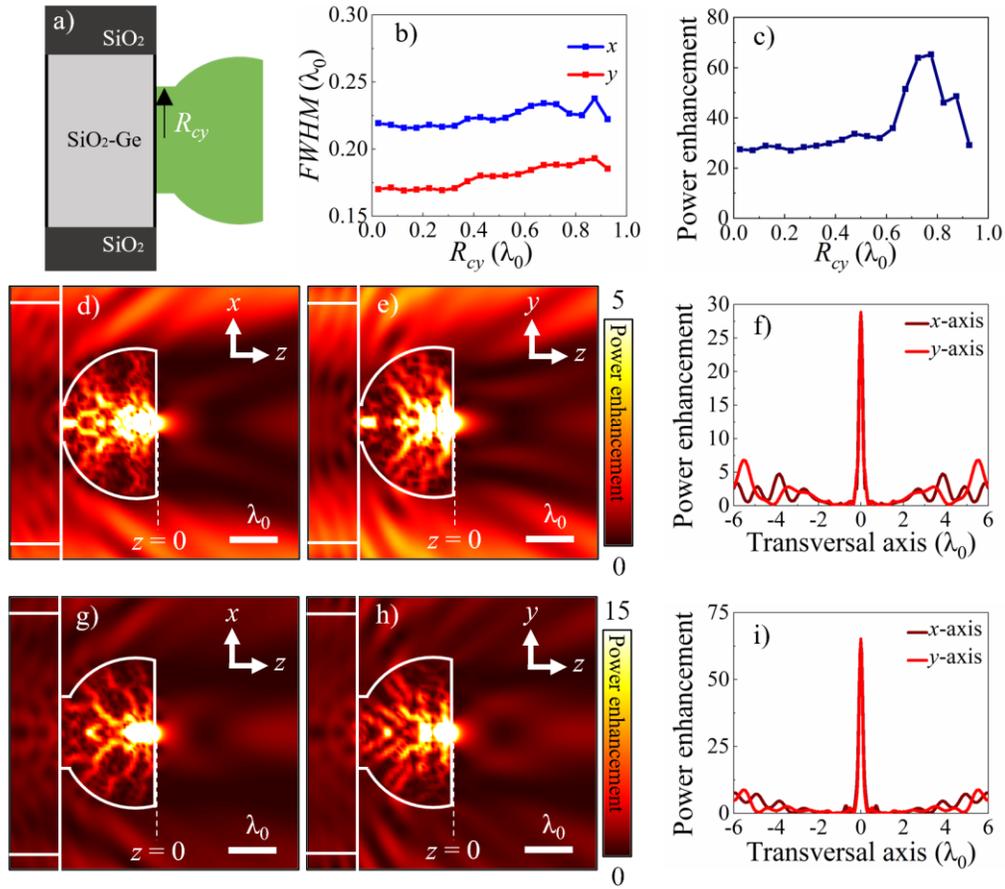

Fig. 6. (a) Shows the schematic of Fig. 4d when the cylinder is added to the back side of the truncated spherical dielectric with radius $R_{cy}$. (b,c) Presents the $FWHM_{x,y}$ and the power enhancement with different cylinder radius $R_{cy}$. (d-f) and (g-i) Show the power enhancement on the *xz*-plane at *y* = 0, *yz*-plane at *x* = 0, and transverse *x*- and *y*-axes at *z* = 0 for $R_{cy} = 0.325\lambda_0$ and $R_{cy} = 0.775\lambda_0$, respectively.

Finally, from an experimental point of view, it is important to study the effect of potential imperfections of the shape of the truncated hemispherical dielectric when it is attached to the optical fiber. To do this, we consider the structure shown in Fig. 6a by adding a cylinder on the back of the structure discussed in Fig. 4d. Here we use the same refractive index of the 3D truncated dielectric sphere. Note that this configuration, in addition to account to fabrication imperfections, it also relaxes the need of using a full hemispherical shape for the 3D dielectric, avoiding the requirement of having a single point of the hemisphere touching the top of the optical fiber (as shown in Fig. 4d). With this setup, we show the effect on the spatial resolution and power enhancement of the generated PNJ when the added cylinder with a radius $R_{cy}$ is changed from $0.025\lambda_0$ to $0.925\lambda_0$ with a step of $0.05\lambda_0$. The spatial resolution on the transversal $x$- and $y$-axes at $z = 0$ is presented in Fig. 6b. As observed, values of *FWHM* of less than $0.25\lambda_0$ and $0.20\lambda_0$, respectively, are obtained for all the designs. For completeness, the numerical results of the power enhancement calculated at $z = 0$ (Fig. 6c) vary from ~27 to 65 when using different values of $R_{cy}$.

By observing the results from Fig. 6b,c it can be seen that the effect of $R_{cy}$ on the spatial resolution of the generated PNJ is almost negligible for values of $R_{cy}$ between $0.025\lambda_0$ and $0.325\lambda_0$. For instance, when a cylinder of $R_{cy} = 0.325\lambda_0$ is added to the back of the 3D truncated dielectric its spatial resolution on the transverse axes, $x$- and $y$-axes at $z = 0$ shown in Fig. 6f, is $0.21\lambda_0$ and $0.17\lambda_0$, respectively (the power enhancement on the $xz$- and $yz$-planes for this value of $R_{cy} = 0.325\lambda_0$ are shown in Fig. 6d,e, respectively) which is the same resolution as the one obtained with the case without adding a cylinder at the back of the 3D truncated dielectric (Fig. 4h,i). For completeness, we present in Fig. 6g-i similar results to those from Fig. 6d-f but considering a value of $R_{cy} = 0.775\lambda_0$ which corresponds to the maximum calculated power enhancement (~65). From these results, a resolution of $0.22\lambda_0$ and $0.18\lambda_0$ is obtained along the $x$- and $y$-axes, respectively, demonstrating that the proposed structure can produce PNJs with high spatial resolution when the back surface of the 3D truncated spherical dielectric is not completely perfect but flat.

## 5. Conclusion

A single-mode optical fiber working at telecommunication wavelengths ($\lambda_0 = 1.55$ μm) has been combined with a high refractive index mesoscale dielectric particle to produce PNJs with a high spatial resolution. Different scenarios were studied using both 2D and 3D cylinders and spheres, respectively. It was shown that, given the high refractive index contrast between the dielectric and the background medium, the PNJ is generated inside the dielectric when using full cylinders/spheres (2D/3D structures). However, it can be shifted near the output surface of the mesoscale dielectrics by truncating their output profile using the Weierstrass formulation for solid immersion lenses. For the 2D designs, a focal spot with a spatial resolution of *FWHM* $= 0.28\lambda_0$ was obtained with a power enhancement of ~4 times the power without using the dielectric cylinder. For the truncated spherical (3D scenario) an improved resolution of $FWHM_x = 0.21\lambda_0$ and $FWHM_y = 0.17\lambda_0$ was obtained with a power enhancement of ~27, considering a design using a dielectric sphere of radius $R = 1.55\lambda_0$ with a truncation parameter of $a \approx 0.41\lambda_0$. It was shown that, while the power enhancement of the PNJ is reduced by increasing losses, the overall spatial resolution is still comparable to the case without losses (with values of $FWHM_{x,y}$ of ~$0.25\lambda_0$ and ~$0.18\lambda_0$, respectively, when including high losses of $\tan \delta = 0.03$. The effect of potential fabrication imperfections of the 3D dielectric was also discussed by considering a cylinder placed at the back of the dielectric to account for a partially flat back surface instead of a perfect spherical profile. It was shown that the spatial resolution is not significantly affected when using the back cylinder with spatial resolutions of the generated PNJs along the $x$- and $y$-axes below $0.25\lambda_0$ and $0.20\lambda_0$, respectively, for different radii of the back cylinder. The proposed combined structure of an optical fiber and a mesoscale dielectric particle to produce PNJs could be used in applications where both a high spatial resolution and high-power enhancement is needed such as in imaging and sensing systems.

## Acknowledgments

V.P.-P. acknowledges support from Newcastle University (Newcastle University Research Fellowship). V.P-P. and J.A.R would like to thank the support from the Engineering and Physical Sciences Research Council (EPSRC) under the EPSRC DTP PhD scheme (EP/R51309X/1).

## Data availability statement

All the data that supports the findings are presented in this manuscript. Specific data can be made available from the corresponding author upon reasonable request.

## Disclosures

The authors declare no conflicts of interest.

## References

1. M. Born and E. Wolf, *Principles Of Optics*, 7th ed. (New York: Cambridge University Press, 1999).
2. Z. Wang and W. Zhang, "Sub-50 nm focusing of a 405 nm laser by hemispherical silicon nanolens," J. Opt. Soc. Am. B **38**(1), 44 (2021).
3. Y. Miao, X. Gao, G. Wang, Y. Li, X. Liu, and G. Sui, "Microsphere-lens coupler with 100 nm lateral resolution accuracy in visible light," Appl. Opt. **59**(20), 6012 (2020).
4. G. Chen, Z. Wu, A. Yu, K. Zhang, J. Wu, L. Dai, Z. Wen, Y. He, Z. Zhang, S. Jiang, C. Wang, and X. Luo, "Planar binary-phase lens for super-oscillatory optical hollow needles," Sci. Rep. **7**(1), 1–10 (2017).
5. S. Legaria, V. Pacheco-Peña, and M. Beruete, "Super-oscillatory metalens at terahertz for enhanced focusing with reduced side lobes," Photonics **5**(4), 10–16 (2018).
6. S. Legaria, J. Teniente, S. Kuznetsov, V. Pacheco-Peña, and M. Beruete, "Highly Efficient Focusing of Terahertz Waves with an Ultrathin Superoscillatory Metalens: Experimental Demonstration," Adv. Photonics Res. **2**(9), 2000165 (2021).
7. A. Poddubny, I. Iorsh, P. Belov, and Y. Kivshar, "Hyperbolic metamaterials," Nat. Photonics **7**(12), 958–967 (2013).
8. V. Pacheco-Peña, B. Orazbayev, V. Torres, M. Beruete, and M. Navarro-Cía, "Ultra-compact planoconcave zoned metallic lens based on the fishnet metamaterial," Appl. Phys. Lett. **103**(18), 1–7 (2013).
9. V. Pacheco-Peña, M. Navarro-Cía, and M. Beruete, "Epsilon-near-zero metalenses operating in the visible," Opt. Laser Technol. **80**, 162–168 (2016).
10. V. Pacheco-Peña, N. Engheta, S. Kuznetsov, A. Gentselev, and M. Beruete, "Experimental Realization of an Epsilon-Near-Zero Graded-Index Metalens at Terahertz Frequencies," Phys. Rev. Appl. **8**(3), 1–10 (2017).
11. J. Yang, I. Ghimire, P. C. Wu, S. Gurung, C. Arndt, D. P. Tsai, and H. W. H. Lee, "Photonic crystal fiber metalens," Nanophotonics **8**(3), 443–449 (2019).
12. J. A. Riley, N. Healy, and V. Pacheco-Peña, "Plasmonic meniscus lenses," Sci. Rep. **12**(1), 1–11 (2022).
13. Z. Wang, W. Guo, L. Li, B. Luk'Yanchuk, A. Khan, Z. Liu, Z. Chen, and M. Hong, "Optical virtual imaging at 50 nm lateral resolution with a white-light nanoscope," Nat. Commun. **2**(1), 1–6 (2011).
14. S. B. Ippolito, B. B. Goldberg, and M. S. Ünlü, "Theoretical analysis of numerical aperture increasing lens microscopy," J. Appl. Phys. **97**(5), 0–12 (2005).
15. W. Zhang and H. Lei, "Fluorescence enhancement based on cooperative effects of a photonic nanojet and plasmon resonance," Nanoscale **12**(12), 6596–6602 (2020).
16. G. Wu and M. Hong, "Optical nano-imaging via microsphere compound lenses working in non-contact mode," Opt. Express **29**(15), 23073 (2021).
17. V. Pacheco-Peña, M. Beruete, I. V. Minin, and O. V. Minin, "Terajets produced by dielectric cuboids," Appl. Phys. Lett. **105**(8), (2014).
18. I. M. Igor Minin and O. M. Oleg Minin, "3D difractive lenses to overcome the 3D Abbe subwavelength difraction limit," Chinese Opt. Lett. **12**(6), 060014–060016 (2014).
19. Z. Chen, A. Taflove, and V. Backman, "Photonic nanojet enhancement of backscattering of light by nanoparticles: a potential novel visible-light ultramicroscopy technique," Opt. Express **12**(7), 1214 (2004).
20. Y. C. Li, H. B. Xin, H. X. Lei, L. L. Liu, Y. Z. Li, Y. Zhang, and B. J. Li, "Manipulation and detection of single nanoparticles and biomolecules by a photonic nanojet," Light Sci. Appl. **5**(12), 1–9 (2016).
21. V. Pacheco-Peña and M. Beruete, "Photonic nanojets with mesoscale high-index dielectric particles," J. Appl. Phys. **125**(8), (2019).
22. I. V. Minin, O. V. Minin, V. Pacheco-Peña, and M. Beruete, "Localized photonic jets from flat, three-dimensional dielectric cuboids in the reflection mode," Opt. Lett. **40**(10), 2329 (2015).
23. A. D. Kiselev and D. O. Plutenko, "Mie scattering of Laguerre-Gaussian beams: Photonic nanojets and

near-field optical vortices," Phys. Rev. A - At. Mol. Opt. Phys. **89**(4), (2014).
24. I. V. Minin, O. V. Minin, and Y. E. Geints, "Localized em and photonic jets from non-spherical and non-symmetrical dielectric mesoscale objects: Brief review," Ann. Phys. **527**(7–8), 491–497 (2015).
25. Y. Zihan and L. Sylvain, "Whispering gallery mode resonance contribution in photonic nanojet simulation," Opt. Express **29**(24), 39249 (2021).
26. X. Li, Z. Chen, A. Taflove, and V. Backman, "Optical analysis of nanoparticles via enhanced backscattering facilitated by 3-D photonic nanojets," Opt. Express **13**(2), 526 (2005).
27. R. Pierron, G. Chabrol, S. Roques, P. Pfeiffer, J.-P. Yehouessi, G. Bouwmans, and S. Lecler, "Large-mode-area optical fiber for photonic nanojet generation," Opt. Lett. **44**(10), 2474 (2019).
28. J. Zelgowski, A. Abdurrochman, F. Mermet, P. Pfeiffer, J. Fontaine, and S. Lecler, "Photonic jet subwavelength etching using a shaped optical fiber tip," Opt. Lett. **41**(9), 2073 (2016).
29. D. Bouaziz, G. Chabrol, A. Guessoum, N.-E. Demagh, and S. Lecler, "Photonic Jet-Shaped Optical Fiber Tips versus Lensed Fibers," Photonics **8**(9), 373 (2021).
30. C. Liu, "Ultra-elongated photonic nanojets generated by a graded-index microellipsoid," Prog. Electromagn. Res. Lett. **37**, 153–165 (2013).
31. M. Salhi and P. G. Evans, "Photonic nanojet as a result of a focused near-field diffraction," J. Opt. Soc. Am. B **36**(4), 1031 (2019).
32. C. Rubio, D. Tarrazó-Serrano, O. V. Minin, A. Uris, and I. V. Minin, "Wavelength-scale gas-filled cuboid acoustic lens with diffraction limited focusing," Results Phys. **12**(February), 1905–1908 (2019).
33. J. H. Lopes, J. P. Leão-Neto, I. V. Minin, O. V. Minin, and G. T. Silva, "A theoretical analysis of acoustic jets," 22nd Int. Congr. Acoust. 1–7 (2016).
34. V. Pacheco-Peña, I. V. Minin, O. V. Minin, and M. Beruete, "Comprehensive analysis of photonic nanojets in 3D dielectric cuboids excited by surface plasmons," Ann. Phys. **528**(9–10), 684–692 (2016).
35. V. Pacheco-Peña, I. V. Minin, O. V. Minin, and M. Beruete, "Increasing surface plasmons propagation via photonic nanojets with periodically spaced 3D dielectric cuboids," Photonics **3**(1), (2016).
36. J. Riley, N. Healy, and V. Pacheco-Pena, "Manipulating Surface Plasmons Propagation Using Ultra-Compact and Non-Dielectric Designs," 2020 14th Int. Congr. Artif. Mater. Nov. Wave Phenomena, Metamaterials 2020 150–152 (2020).
37. D. Ju, H. Pei, Y. Jiang, and X. Sun, "Controllable and enhanced nanojet effects excited by surface plasmon polariton," Appl. Phys. Lett. **102**(17), (2013).
38. S. Lee and L. Li, "Rapid super-resolution imaging of sub-surface nanostructures beyond diffraction limit by high refractive index microsphere optical nanoscopy," Opt. Commun. **334**, 253–257 (2015).
39. Y. Li, H. Xin, X. Liu, Y. Zhang, H. Lei, and B. Li, "Trapping and Detection of Nanoparticles and Cells Using a Parallel Photonic Nanojet Array," ACS Nano **10**(6), 5800–5808 (2016).
40. V. R. Dantham, P. B. Bisht, and C. K. R. Namboodiri, "Enhancement of Raman scattering by two orders of magnitude using photonic nanojet of a microsphere," J. Appl. Phys. **109**(10), (2011).
41. J. Wenger and H. Rigneault, "Photonic methods to enhance fluorescence correlation spectroscopy and single molecule fluorescence detection," Int. J. Mol. Sci. **11**(1), 206–221 (2010).
42. V. N. Astratov, A. Darafsheh, M. D. Kerr, K. W. Allen, N. M. Fried, A. N. Antoszyk, and H. S. Ying, "Photonic nanojets for laser surgery," SPIE Newsroom (2010).
43. B. S. Luk'yanchuk, R. Paniagua-Domínguez, I. Minin, O. Minin, and Z. Wang, "Refractive index less than two: photonic nanojets yesterday, today and tomorrow [Invited]," Opt. Mater. Express **7**(6), 1820 (2017).
44. I. V Minin, O. V Minin, V. Pacheco-Peña, and M. Beruete, "Subwavelength, standing-wave optical trap based on photonic jets," Quantum Electron. **46**(6), 555–557 (2016).
45. A. Heifetz, K. Huang, A. V. Sahakian, X. Li, A. Taflove, and V. Backman, "Experimental confirmation of backscattering enhancement induced by a photonic jet," Appl. Phys. Lett. **89**(22), 1–4 (2006).
46. V. Pacheco-Peña, M. Beruete, I. V. Minin, and O. V. Minin, "Multifrequency focusing and wide angular scanning of terajets," Opt. Lett. **40**(2), 245 (2015).
47. A. Heifetz, S. C. Kong, A. V. Sahakian, A. Taflove, and V. Backman, "Photonic nanojets," J. Comput. Theor. Nanosci. **6**(9), 1979–1992 (2009).
48. B. Wei, S. Gong, R. Li, I. V. Minin, O. V. Minin, and L. Lin, "Optical Force on a Metal Nanorod Exerted by a Photonic Jet," Nanomaterials **12**(2), (2022).
49. V. Gašparić, T. G. Mayerhöfer, D. Zopf, D. Ristić, J. Popp, and M. Ivanda, "To generate a photonic nanojet outside a high refractive index microsphere illuminated by a Gaussian beam," Opt. Lett. **47**(10), 2534 (2022).
50. S. M. Mansfield and G. S. Kino, "Solid immersion microscope," Appl. Phys. Lett. **57**(24), 2615–2616 (1990).
51. P. Ferrand, J. Wenger, A. Devilez, M. Pianta, B. Stout, N. Bonod, E. Popov, and H. Rigneault, "Direct imaging of photonic nanojets," Opt. Express **16**(10), 6930 (2008).
52. S. C. Kong, A. V. Sahakian, A. Heifetz, A. Taflove, and V. Backman, "Robust detection of deeply subwavelength pits in simulated optical data-storage disks using photonic jets," Appl. Phys. Lett. **92**(21), 1–4 (2008).
53. W. Aljuaid, N. Healy, and V. Pacheco-Pena, "High-refractive-index Nanosparticles on Optical Fibres for High-Resolution Lensing Applications," 2020 14th Int. Congr. Artif. Mater. Nov. Wave Phenomena, Metamaterials 2020 228–230 (2020).

54. Z. Li, J. Yang, S. Liu, X. Jiang, H. Wang, X. Hu, S. Xue, S. He, and X. Xing, "High throughput trapping and arrangement of biological cells using self-assembled optical tweezer," Opt. Express **26**(26), 34665 (2018).
55. P. Sharma, R. K. Arora, S. Pardeshi, and M. Singh, "Fibre Optic Communications : An Overview," Int. J. Emerg. Technol. Adv. Eng. **3**(5), 474–479 (2013).
56. A. Manickavasagan and H. Jayasuriya, *Imaging with Electromagnetic Spectrum* (Springer Heidelberg New York Dordrecht London, 2014).
57. A. N. K. Reddy and D. Karuna Sagar, "Half-Width at half-Maximum, full-Width at half-Maximum analysis for resolution of asymmetrically apodized optical systems with slit apertures," Pramana - J. Phys. **84**(1), 117–126 (2015).